\documentclass[sn-mathphys]{sn-jnl}

\jyear{2022}%

\theoremstyle{thmstyleone}%
\theoremstyle{thmstyletwo}%

\theoremstyle{thmstylethree}%

\begin{document}

\title{Bremsstrahlung as a probe of baryon stopping in heavy-ion collisions}

\author{\fnm{Sigurd} \sur{Nese}}\email{sigun@uio.no}

\author{\fnm{Joakim} \sur{Nystrand}}\email{Joakim.Nystrand@uib.no}

\affil{\orgdiv{Department of Physics and Technology}, \orgname{University of Bergen},
  \city{Bergen},
  \country{Norway}}

\abstract{In collisions between heavy ions at ultra-relativistic energies the participating protons lose energy, which is
  converted into new particles. As the protons slow down, they emit bremsstrahlung radiation. The yield
  and angular distribution of the emitted radiation are sensitive probes of how much energy the incoming protons
  have lost. In this paper, the spectrum of bremsstrahlung radiation is calculated for different stopping scenarios, and
  the results are compared with the expected yield of photons from hadronic interactions.}




\maketitle

\section{Introduction}\label{sec1}

In collisions between heavy ions at relativistiv energies there is convincing evidence that a new state of matter,
a quark-gluon plasma, is formed~\cite{Muller:2012zq}. In this state, the quarks and gluons are no longer confined
to nucleons but can
move freely over distances large compared with the size of a single nucleon. The energy density in the quark gluon
plasma formed in Pb+Pb collisions at the LHC has been estimated from measurements of the total transverse energy
at midrapidity to be on the order of 12-14 GeV/fm$^3$ at a time of 1 fm/c after the
collision~\cite{CMS:2012krf,ALICE:2016igk}. This is far above the densitites of 0.2-0.5 GeV/fm$^3$ lattice QCD
calculations find are required for deconfinement~\cite{HotQCD:2014kol}. 

The energy deposited in the quark-gluon plasma comes from the energy lost by the incoming nuclei, and 
one of the most fundamental question one can address in the study of
high energy heavy-ion collisions is therefore how much energy the incoming baryons lose.   
This is usually referred to as the amount of baryon stopping. One can have scenarios ranging from
complete stopping, where the incoming baryons lose all their energy,
to full transparency, where the baryons lose no or very little energy.
Full stopping would imply that all baryons end up close to midrapidity, whereas full transparency
would leave the baryons near beam rapidity. 
Since baryon number is conserved, 
the fate of the baryons in the colliding nuclei can be determined
from the rapidity distribution of net baryons, that is $dn_B/dy - dn_{\overline{B}}/dy$.
For experimental reasons, one is often restricted to study the net proton
rather than the net baryon distributions. 

Results from the Relativistic Heavy-Ion Collider (RHIC)~\cite{STAR:2017sal,BRAHMS:2009wlg,BRAHMS:2003wwg}
and fixed target experiments at the CERN SPS~\cite{NA49:1998gaz} show that the amount of stopping decreases with increasing
collision energy in the range $\sqrt{s_{NN}} = 7 - 200$~GeV. There have been attempts to
explain the energy loss in this energy range from hadron transport models~\cite{Mohs:2019iee} and
models based on the Color Glass Condensate~\cite{Li:2018ini,McLerran:2018avb}.

At the CERN SPS and RHIC, identified protons and anti-protons could be measured down to low transverse momenta $p_T \sim 0$
over a wide rapidity range, and the net-proton rapidity distributions could thus be determined. Such measurements were
performed by the NA49~\cite{NA49:1998gaz} and BRAHMS~\cite{BRAHMS:2009wlg,BRAHMS:2003wwg} experiments.
At the LHC, the situation
is different. The only experipment which has measured identified protons and anti-protons at low $p_T$ is ALICE.
The results have shown that in the central rapidity region $\rvert y \rvert \leq 0.5$ there are no net
protons~\cite{ALICE:2012ovd,ALICE:2013mez,ALICE:2019hno}.
But beyond that, there are no experimental constraints on how the net-protons are distributed. 

To improve this situation, we propose to use the bremsstrahlung photons emitted when the nuclei slow down. This idea
was first suggested before the start of the relativistic heavy-ion programs at CERN and  
RHIC~\cite{Kapusta:1977zb,Bjorken:1984sp}. Before the start-up of RHIC, several studies
were made where this process was considered~\cite{Dumitru:1993ph,Jeon:1998tq,Kapusta:1999hb}. These also included a
proposal to build a dedicated detector to study this radiation~\cite{Jeon:1998tq}, but those plans were never
realized. Recently, the idea was brought up again in the context of the LHC~\cite{Park:2021ljg}. 

The previous studies mentioned above all use a similar, semi-classical approach to calculate the bremsstrahlung
spectrum, based on the description in~\cite{Jackson1975}. Our calculations will follow the same path. We will, 
however, implement improved stopping scenarios which are either based on model calculations or phenomenological
and consistent with existing data. 
The stopping scenarios considered in~\cite{Park:2021ljg} are simplified and do not take into account the fact
that the central region ($\lvert y \rvert \leq 0.5$) at the LHC is baryon free. This was, however, recently 
followed up by a study of phenomenological stopping scenarios where the central region is almost
baryon free~\cite{Park:2022kcs}.

We will also, for the first time, make a
detailed estimate of the background from hadronically
produced photons, primarily from the decay of $\pi^0$ mesons. This background is obtained from simulations with
PYTHIA 8.3~\cite{Bierlich:2022pfr}. 
The goal is to determine in what regions of phase space one can expect bremsstrahlung photons to provide a realistic
measure of the nuclear stopping. One can, in addition to photons from hadronic interactions, also expect a large
background from secondary photons produced in the detector material. The latter is specific to a certain experiment
and its material budget and is thus beyond the scope of this paper.

\section{The bremsstrahlung spectrum}\label{sec2}

The energy radiated per solid angle from a current ${\bf J}({\bf r},t)$ is given by~\cite{Jackson1975}
\begin{equation}
  \frac{d^2I}{d \omega d \Omega} = \frac{\omega^2}{4 \pi^2} \Bigl\lvert \int \int \overline{n} \times (\overline{n} \times {\bf J}({\bf r},t))
  e^{i \omega (t - \overline{n} \cdot {\bf r}(t))} dt d^3{\bf r} \Bigr\rvert^2 \, .
\label{dIJackson}
\end{equation}
The vector $\overline{n}$ is a unit vector in the direction of the photon, and here and throughout we use units where $\hbar = c = 1$. 
We choose coordinates where the incoming beams move along the z-axis and the photon is emitted in the xz-plane with angle $\theta$, giving
$\overline{n} = (\sin(\theta),0,\cos(\theta))$.

In the center of mass, the incoming nuclei are Lorentz contracted in the longitudinal direction to a size $\sim R/\gamma$, where $R$ is the
nuclear radius and $\gamma$ the Lorentz factor of the beam. For a lead nucleus at the LHC, this corresponds to a longitudinal size of about
0.003 fm. It is thus justified to ignore the longitudinal relative to the transverse extension and write the currents for the
incoming nuclei as 
\begin{equation}
\begin{array}{l}
  {\bf J}_{+} ({\bf r},t) = + v_0 \frac{e}{\sqrt{4 \pi \epsilon_0}} \sigma(r_{\perp}) \delta(z - v_0 t) \, \theta(-t) \hat{{\bf z}} \\ \\
  {\bf J}_{-} ({\bf r},t) = - v_0 \frac{e}{\sqrt{4 \pi \epsilon_0}} \sigma(r_{\perp}) \delta(z + v_0 t) \, \theta(-t) \hat{{\bf z}} \, .\\
\end{array}
\end{equation}
Here, $\sigma$ is the nuclear electric charge density in the transverse plane, $v_0$ the velocity,
$v_0 = \tanh(y_b)$, where $y_b$ is the beam rapidity, and $\theta(t)$ the Heaviside step function.
The charge density is normalized to $\int \sigma d^2{\bf r}_{\perp} = Z$, where $Z$ is
the number of protons in the beam nucleus. 
The outgoing protons will have a distribution in the transverse plane, which we assume is the same as for the
incoming particles, and a distribution in velocities, which may or may not depend on the position in the transverse plane.
Writing the velocity in terms of the rapidity, $v(y) = \tanh(y)$, and the corresponding density as $\rho(y, r_{\perp})$
one gets for the outgoing current 
\begin{equation}
{\bf J}_f ({\bf x},t) =  \sigma(r_{\perp})  \frac{e}{\sqrt{4 \pi \epsilon_0}} \int_{-\infty}^{+ \infty} \rho(r_{\perp}, y) v(y) \delta(z - v(y) t) dy \, \theta(t) \hat{{\bf z}} \, .
\end{equation}
The density of the outgoing protons is normalized to
\begin{equation}
\int_{-\infty}^{+ \infty} \rho(r_{\perp}, y) dy = 2 \, ,
\end{equation}
since it includes the contribution from both incoming nuclei. With the total current
${\bf J} = {\bf J}_+ +  {\bf J}_- +  {\bf J}_f$, the radiated energy becomes 
\begin{equation}
  \begin{array}{l}
  \frac{d^2I}{d \omega d \Omega} = \frac{\omega^2}{4 \pi^2} \Bigl\lvert
  \int \int \overline{n} \times (\overline{n} \times {\bf J}_+({\bf r},t)) e^{i \omega (t - \overline{n} \cdot {\bf r}(t))} dt d^3{\bf r} \, +  \\ \\
  \int \int \overline{n} \times (\overline{n} \times {\bf J}_-({\bf r},t)) e^{i \omega (t - \overline{n} \cdot {\bf r}(t))} dt d^3{\bf r} \, + \\ \\
  \int \int \overline{n} \times (\overline{n} \times {\bf J}_f({\bf r},t)) e^{i \omega (t - \overline{n} \cdot {\bf r}(t))} dt d^3{\bf r} 
  \Bigr\rvert^2 \, .
  \end{array}
\end{equation}
Integrating by parts in the integral over time one obtains
\begin{equation}
  \begin{array}{l}
  \frac{d^2I}{d \omega d \Omega} = \frac{\omega^2}{4 \pi^2} \Bigl\lvert
  \frac{1}{i \omega} \int \left[ \int \frac{d}{dt} \left[ \frac{\overline{n} \times (\overline{n} \times {\bf J}_+({\bf r},t))}{1 - \overline{n} \cdot \dot{\bf r}(t)} \right] 
      e^{i \omega (t - \overline{n} \cdot {\bf r}(t))} dt \right] d^3{\bf r} \, +  \\ \\
  \frac{1}{i \omega} \int \left[ \int \frac{d}{dt} \left[ \frac{\overline{n} \times (\overline{n} \times {\bf J}_-({\bf r},t))}{1 - \overline{n} \cdot \dot{\bf r}(t)} \right] 
      e^{i \omega (t - \overline{n} \cdot {\bf r}(t))} dt \right] d^3{\bf r} \, + \\ \\
      \frac{1}{i \omega} \int \left[ \int \frac{d}{dt} \left[ \frac{\overline{n} \times (\overline{n} \times {\bf J}_f({\bf r},t))}{1 - \overline{n} \cdot \dot{\bf r}(t)} \right] 
      e^{i \omega (t - \overline{n} \cdot {\bf r}(t))} dt \right] d^3{\bf r} \,
  \Bigr\rvert^2 \, .
  \end{array}
  \label{partint}
\end{equation}

The time, $\Delta t$, and longtudinal distance over which the protons are slowed down can be expected to be small compared with
the transverse size, $\Delta t << R$.  
For low energy photons ($\omega << 1/\Delta t$), it is therefore justified to neglect
the time and longitudinal components in the phase factor and make the assupmtion
\begin{equation}
e^{i \omega (t - \overline{n} \cdot {\bf r})} \approx e^{-i \omega x \sin(\theta)} \, .
\end{equation}
With these assumptions the integrals over time and z in \eqref{partint} can be performed, leading to
\begin{equation}
  \begin{array}{l}
    \frac{d^2I}{d \omega d \Omega} = 
    \frac{\alpha}{4 \pi^2} \sin^2(\theta) \Bigl\lvert
    \int \sigma(r_{\perp}) e^{i \omega x \sin(\theta)}
    \left[ \int \frac{v(y)}{1 - v(y) \cos(\theta)} \rho(y, r_{\perp}) dy - \right. \\ \\
    \left. \frac{2 v_0^2 \cos(\theta)}{1 -
    v_0^2 \cos^2(\theta)} \right] d^2r_{\perp} 
    \Bigr\rvert^2 \, .
  \end{array}
\end{equation}

Here, $\alpha = e^2/4 \pi \epsilon_0$ is the fine structure constant.
This result is in agreement with~\cite{Jeon:1998tq}. To do the integral
over the transverse dimensions, one has to know the function $\rho(y, r_{\perp})$. It is
conceivable that there is a dependence of the rapidity loss of the protons on the transverse coordinate;
protons close to the center of the nuclei
can be expected to lose more energy than those on the periphery. Previous studies have, however, found that
the dependence of $\rho(y, r_{\perp})$ on the transverse
position has only a minor effect on the spectrum of bremsstrahlung photons~\cite{Jeon:1998tq,Kapusta:1999hb}.
Moreover, the models we will use for the rapidity loss of the protons provide the average loss, independent of position
in the transverse plane. 
We will therefore ignore the dependence on $ r_{\perp}$ here
and assume $\rho(y, r_{\perp}) = \rho(y)$. The integral over $r_{\perp}$ can then be performed and the spectrum of
emitted photons can be written 
\begin{equation}
  \begin{array}{l}
  \frac{dN_{\gamma}}{d \omega d \Omega} = \frac{\alpha Z^2}{4 \pi^2 \omega} \sin^2(\theta) \lvert F(\omega \sin(\theta)) \rvert^2
  \left[ \int \frac{v(y) \rho(y)}{1 - v(y) \cos(\theta)} dy - \right. \\ \\ \left.
    \frac{2 v_0^2 \cos(\theta)}{1 - v_0^2 \cos^2(\theta)}
    \right]^2 \, .
  \end{array}
\label{photonspectrum}
\end{equation}
Here, $F(Q)$ is the nuclear form factor obtained from a Fourier transform of the nuclear charge distribution, $\rho_A$:
\begin{equation}
  F(Q) = \frac{1}{Z} \int \rho_A({\bf r}) e^{- i {\bf q} \cdot {\bf r}} d^3{\bf r} \approx
  \frac{1}{Z} \int \sigma(r_{\perp}) e^{-i \omega x \sin(\theta)} d^2r_{\perp} \, .
\end{equation}
We use a form factor 
\begin{equation}
F(Q) = \frac{4 \pi \rho_0}{A Q^3} \left( \sin(Q R_A) - Q R_A \cos(Q R_A) \right) \left[ \frac{1}{1 + a^2 Q^2} \right] \, ,
\end{equation}
where $R_A = 6.62$~fm, $\rho_0 = 0.161$~fm$^{-3}$, and $a = 0.70$~fm for a Pb nucleus. This parameterization has been shown
to reproduce the Fourier transform of a Woods-Saxon distribution in configuration space very well~\cite{Klein:1999qj}. 

One can note that for low energy photons and small emission angles, $\omega \sin(\theta) << 1/R$, the Form Factor in
Eq.~\ref{photonspectrum} is approximately 1. The energy and angular dependencies then factorize, with the energy dependence
given by $1/\omega$ and the angular dependence by 
\begin{equation}
  \frac{dN_{\gamma}}{d \theta} \propto \sin^3(\theta) 
  \left[ \int \frac{v(y) \rho(y)}{1 - v(y) \cos(\theta)} dy - \frac{2 v_0^2 \cos(\theta)}{1 - v_0^2 \cos^2(\theta)} \right]^2 \, .
\end{equation}

To proceed with the calculations one has to define the function $\rho(y)$, the rapidity distribution of the net-protons
in the final state. We consider 3 scenarios for central Pb+Pb collisions at the LHC.

\begin{enumerate}

\item The net proton distribution as given by PYTHIA 8.3~\cite{Bierlich:2022pfr}. Heavy-ion collisions have been implemented
in PYTHIA through the Angantyr model~\cite{Bierlich:2018xfw}. The scaling from proton-proton collisions is based on the
Glauber model and the original PYTHIA framework is used to describe the individual nucleon–-nucleon sub-collisions.  It thus
extrapolates the dynamics of pp collisions to heavy-ion collisions, without introducing any collective effects between the
nucleon-nucleon collisions. It does reproduce the measured charged particle pseudorapidity distributions in Pb+Pb collisions
at the LHC. 

\item The net proton distribution as given by the hadron transport model SMASH-2.2~\cite{Weil:2016zrk,Mohs:2019iee}. The model
combines a string
model, where the colliding hadrons are excited to strings which fragment, with elastic and inelastic interaction between
hadrons in the later stages of the collisions. This approach leads to a considerably larger amount of stopping compared
with PYTHIA. In fact, the model predicts a non-zero number of net-protons at midrapidity. We use this result for the
current calculation anyway, since it represents a valid result of the model at LHC energies [H.~Elfner, private communication]. 

\item A phenomenological model where the central region, $\lvert y \rvert \leq 0.5$, contains no net protons, but where the protons
have a considerable shift away from beam rapidity. This is modelled by the sum of two skewed Gaussians.

\end{enumerate}

\begin{figure}[h]%
\centering
\includegraphics[width=0.75\textwidth]{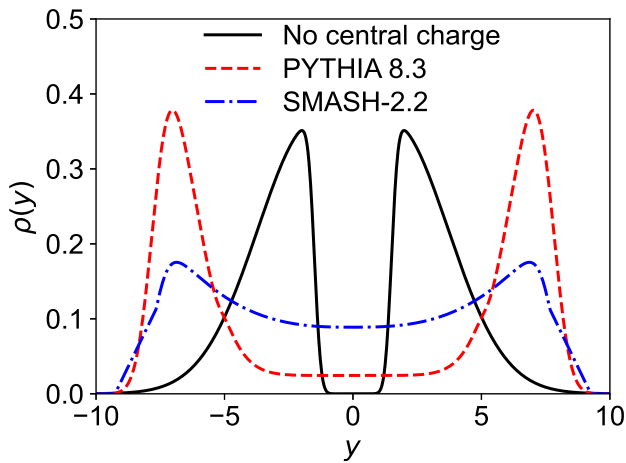}
\caption{Net-proton rapidity distributions for the 3 scenarios described in the text.}\label{rho}
\end{figure}

The net-proton rapidity distributions, $\rho(y)$, for these 3 scenarios are shown in Fig.~\ref{rho}. Of the three models, Pythia
clearly shows the least amount of stopping. While SMASH-2.2 exhibits a non-zero yield of net-protons at midrapidity, the protons
are on average not shifted so much toward $y = 0$ as in scenario 3.

\section{Results}\label{sec3}

\begin{figure}[h]%
\centering
\includegraphics[width=0.6\textwidth]{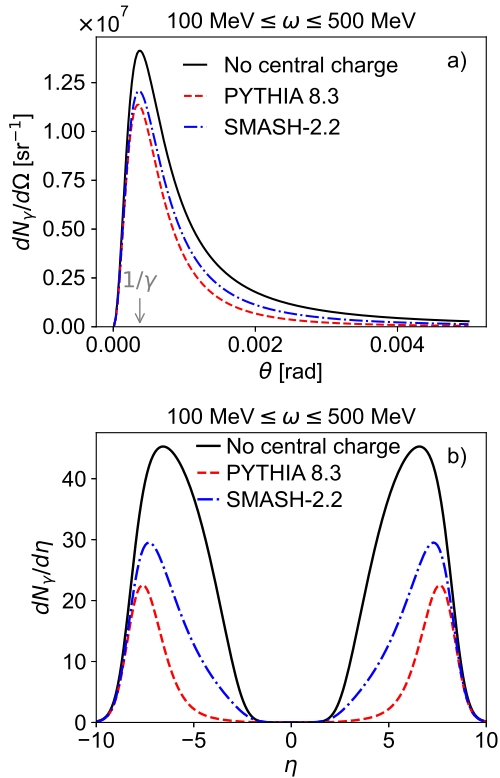} 
\caption{a) The angular distribution of bremsstrahlung photons with energies between $0.1 \leq \omega \leq 0.5$~GeV for
  the 3 scenarios. b) The pseudorapidity distributions, $dN_{\gamma}/d \eta$, of bremsstrahlung photons within the same
  energy range.}\label{global}
\end{figure}

The angular distributions of bremsstrahlung photons in the forward direction, integrated over the energy range
$0.1 \leq \omega \leq 0.5$~GeV, from the 3 scenarios are shown in Fig.~\ref{global} a). The distributions are peaked close
to $1/\gamma$, as expected, and the yields increase with an increasing amount of stopping. The curves also exhibit
different angular dependencies, the scenarios with more stopping having more persistent tails toward larger angles.
Thus, the models are differentiated in both total photon yield and angular dependence.

To facilitate a comparison with experimental acceptances, which are usually defined in terms of pseudorapidity,
$\eta = - \ln(\tan(\theta/2))$, the angular distribution in Eq.~\ref{photonspectrum} can be rewritten as
\begin{equation}
\frac{dN_{\gamma}}{d \omega d \eta d \phi} = \frac{dN_{\gamma}}{d \omega d \Omega} \sin^2(\theta) \, .
\label{dndetaformula}
\end{equation}
The pseudorapidity distribution, integrated over azimuthal angle and energy interval $0.1 \leq \omega \leq 0.5$~GeV, is shown in
Fig.~\ref{global} b). The difference between the spectra is more pronounced in the pseudorapidity distribution than
in the angluar distribution. This is due to the factor $\sin^2\theta=1/\cosh^2(\eta)$ which varies rapidly for
large $\lvert \eta \rvert$. Covering the entire range of emission angles, the figure illustrates the difference in total number of
radiated photons between the stopping scenarios. Furthermore, while the peaks of all three spectra lie at large
pseudorapidities, the scenarios give significantly differing photon yields at lower $\eta$ as well, making such
scenarios potentially discernable within experimental acceptances.

\begin{figure}[!th]%
\centering
\includegraphics[width=0.6\textwidth]{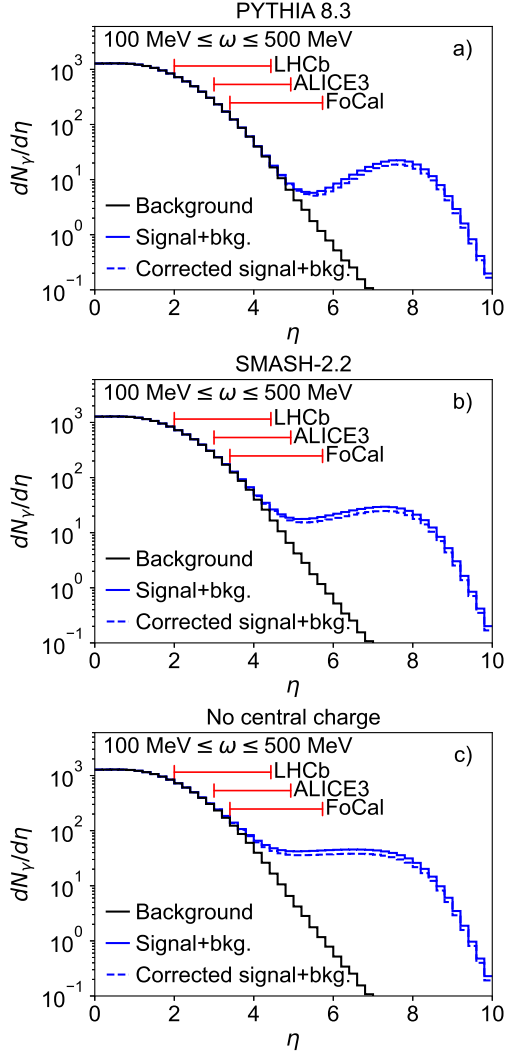}
\caption{The pseudorapidity distributions for photons with $0.1 \leq \omega \leq 0.5$~GeV integrated over the azimuthal angle.
  The black histogram shows the background from hadronically produced photons. The solid blue histogram shows the sum of the
  photons from bremsstrahlung radiation and hadronic production. The correction applied to the bremsstrahlung spectrum
  to obtain the dashed blue histogram is described in the text.}\label{dndeta}
\end{figure}

To put these numbers in context, we compare them with the photon yield from the 5\% most central Pb+Pb collisions from
PYTHIA 8.3 in Fig.~\ref{dndeta}. These hadronically produced photons, most of which come from the
deacy $\pi^0 \rightarrow \gamma + \gamma$, constitute a background to the bremsstrahlung photons we
are considering here. The background yield is shown by the black histograms in the figure. The sum of the
yield of the background and bremsstrahlung photons is shown by the solid, blue histograms for scenario 1-3 in
Fig.~\ref{dndeta} a), b), and c), respectively. The number of bremsstrahlung photons is calculated from
Eq.~\ref{dndetaformula}, integrated over azimuthal angle and photon energy $0.1 \leq \omega \leq 0.5$~GeV.

The bremsstrahlung calculations above assume that all protons participate in the collision. 
This will not be the case  
for collisions within a finite impact parameter range. We therefore calculate a correction factor
$(\langle N_{p,\,\mathrm{part}}\rangle/2Z)^2$, where $\langle N_{p,\,\mathrm{part}}\rangle = 151$ is the average number of
participating protons in the 5\% most central collisions in Pythia. The result of applying this correction factor to the yield of
bremsstrahlung photons is shown by the dashed, blue histograms in the figure. 

From the figure one can see that at low pseudorapidities, the background completely dominates. In the very forward
direction, however, the background falls off quickly while the bremsstrahlung peak emerges. The yield of bremsstrahlung
photons differ widely between the different scenarios, which emphasizes that this is indeed a very sensitive probe of
the amount of nuclear stopping. The shapes of the pseudorapidity distributions are also quite different between the
scenarios. This means that the limit in pseudorapidity where one can expect a significant signal over background is
lower the larger amount of stopping one has. 

As mentioned, a detailed discussion of in which experiments one might extract a bremsstrahlung signal is beyond the
scope of this paper. We nevertheless indicate in Fig.~\ref{dndeta} the experimental acceptances of the current and
future experiments where it might be possible. The existing LHCb experiment has an electromagnetic calorimeter
coverage between $2.0 \leq \eta \leq 4.5$~\cite{LHCb:2014set}. We include it here, although it might not be able to
reach low enough photon energies at large pseudorapidities~\cite{Klusek-Gawenda:2019ijn},~[R.~McNulty, private communication].
During the Next Long Shutdown at the LHC (2026 - 2029) it
is foreseen to install a forward calorimeter (FoCal) in the ALICE experiment~\cite{ALICE:2020mso}. It will consist
of a high resolution electromagnetic and hadronic calorimeter covering $3.4 \leq \eta \leq 5.8$. 
Finally, beyond LHC Run 4
there are plans to upgrade the ALICE experiment to ALICE-3~\cite{ALICE:2803563}. The current design of ALICE-3 includes
a Forward Conversion Tracker, which should have the possibility to measure photons with energies down to or below 100~MeV in
the pseudorapidity range $3.0 \leq \eta \leq 5.0$. The pseudorapidity coverages of these detectors are shown by the red
lines in Fig.~\ref{dndeta}. 

From the figure one can see that for scenario 3, ``No central charge'', there is a visible excess over the hadronic
background within the pseudorapidity coverage of all the detectors mentioned above. For scenario 1, ``PYTHIA 8.3'',
the situation is less favorable and one would have to go to the most forward regions of ALICE-3 and FoCal to find a
good signal to background ratio. 

\begin{figure}[!th]%
\centering
\includegraphics[width=0.6\textwidth]{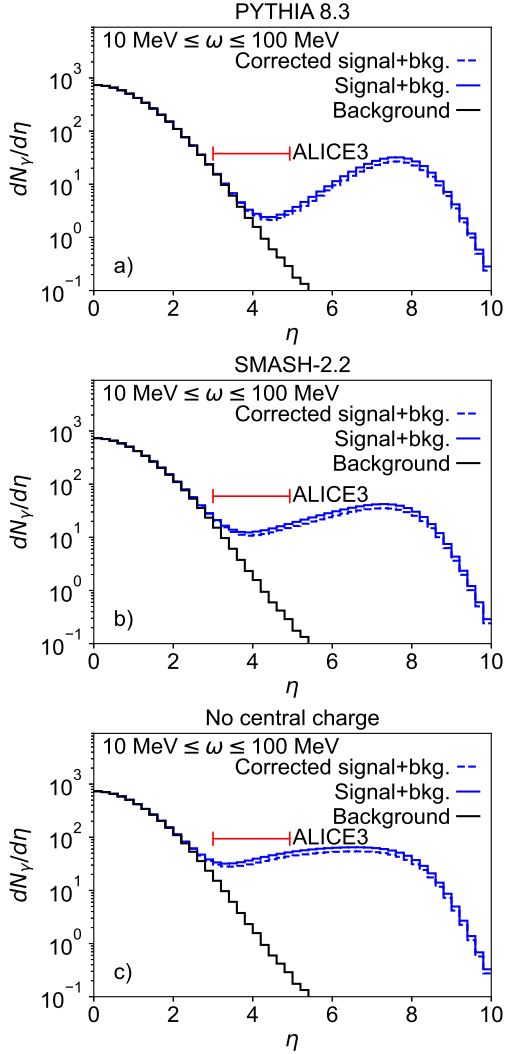}
\caption{Same as Fig.~\ref{dndeta} but for photons in the energy range $0.01 \leq \omega \leq 0.1$~GeV.}\label{dndeta2}
\end{figure}

Since the Photon Conversion Tracker in ALICE-3 aims at measuring photons with energies below 100~MeV, we also include
a plot of the photon pseudorapidity distributions for the photon energy range $0.01 \leq \omega \leq 0.1$~GeV in Fig.~\ref{dndeta2}.
In this energy range, there is a significant excess inside the ALICE-3 acceptance for all stopping scenarios. 

\begin{figure}[!th]%
\centering
\includegraphics[width=0.6\textwidth]{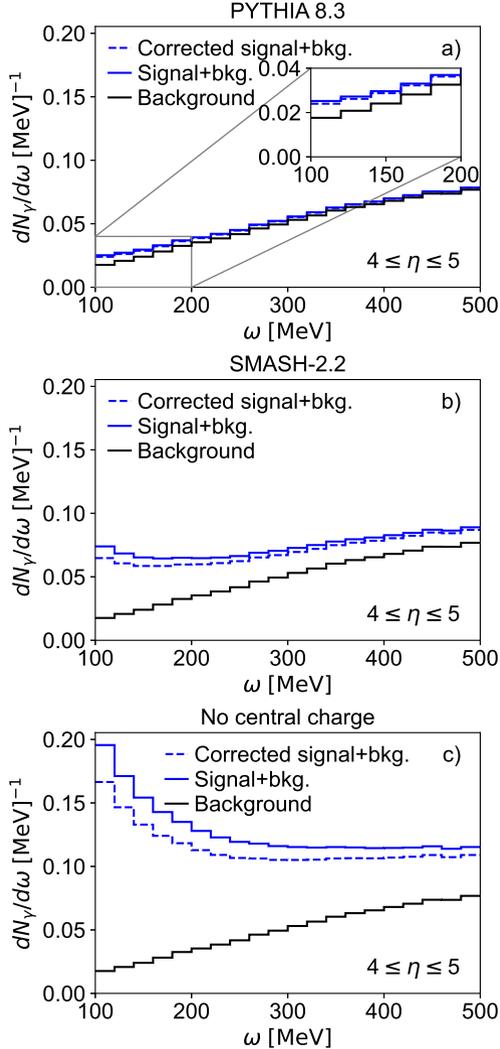}
\caption{The energy distributions for photons with $4.0 \leq \eta \leq 5.0$ integrated over the azimuthal angle.
  The black histogram shows the background from hadronically produced photons. The solid blue histogram shows
  the sum of the photons from bremsstrahlung radiation and hadronic production. The correction applied to the
  bremsstrahlung spectrum to obtain the dashed blue histogram is described in the text. The inset in a) shows
  the low energy region.}\label{dndomega}
\end{figure}

In addition to being peaked in the forward direction, the bremsstrahlung spectrum increases rapidly with decreasing
photon energy, approximately as $1/\omega$,
as was mentioned above. This is contrary to the background from hadronically produced photons, which decrease with
decreasing $\omega$ in the energy range considered here. 
To illustrate this, we plot the energy spectrum integrated over azimuthal angle and pseudorapidity range
$4 \leq \eta \leq 5$ in Fig.~\ref{dndomega}. As in Figs.~\ref{dndeta} and \ref{dndeta2},
the background is given by the black histograms, and the uncorrected and corrected
signal plus background by the solid and dashed blue histograms, respectively. As for the pseudorapidity distributions,
the energy below which one can expect a significant signal over background is highly dependent on the stopping scenario.
Also the yield is strongly dependent on the stopping scenario.

The inset in Fig.~\ref{dndomega} a) shows the low energy region, and it emphasizes that if one can go to low enough photon
energies, a signal will be visible also in scenarios with a small amount of stopping. One should keep in mind that
it might be possible to extract a bremsstrahlung signal, even with a rather low signal to background ratio, by subtracting
the hadronic background. The hadronic background should be well constrained from measurements of charged
and neutral particle spectra. 

\section{Summary}\label{sec4}

To conclude, we have shown that even with realistic stopping scenarios the bremsstrahlung spectra show a strong
sensitivity to the amount of nuclear stopping. Comparisons with Pythia show that a significant signal over the hadronic
background is obtained in the range $\eta \gtrsim 4-5$ and $\omega \lesssim 300 - 500$~MeV. Again, the exact
limits depend on the amount of stopping one has.

Considering the importance of determining the amount of stopping in heavy-ion collisions at the LHC, and given that no
alternative methods are available, we believe the possibility to use bremsstrahlung photons should be considered
seriously. Hopefully this paper can help in the design of future detectors to accomplish such a measurement. 

\bmhead{Acknowledgments}

We thank Hannah Elfner for providing the net proton distributions from SMASH-2.2. This work is 
supported by the Norwegian Research Council.

\newpage
\bibliography{bremsstrahlung}


\begin{thebibliography}{30}
\ifx \bisbn   \undefined \def \bisbn  #1{ISBN #1}\fi
\ifx \binits  \undefined \def \binits#1{#1}\fi
\ifx \bauthor  \undefined \def \bauthor#1{#1}\fi
\ifx \batitle  \undefined \def \batitle#1{#1}\fi
\ifx \bjtitle  \undefined \def \bjtitle#1{#1}\fi
\ifx \bvolume  \undefined \def \bvolume#1{\textbf{#1}}\fi
\ifx \byear  \undefined \def \byear#1{#1}\fi
\ifx \bissue  \undefined \def \bissue#1{#1}\fi
\ifx \bfpage  \undefined \def \bfpage#1{#1}\fi
\ifx \blpage  \undefined \def \blpage #1{#1}\fi
\ifx \burl  \undefined \def \burl#1{\textsf{#1}}\fi
\ifx \doiurl  \undefined \def \doiurl#1{\url{https://doi.org/#1}}\fi
\ifx \betal  \undefined \def \betal{\textit{et al.}}\fi
\ifx \binstitute  \undefined \def \binstitute#1{#1}\fi
\ifx \binstitutionaled  \undefined \def \binstitutionaled#1{#1}\fi
\ifx \bctitle  \undefined \def \bctitle#1{#1}\fi
\ifx \beditor  \undefined \def \beditor#1{#1}\fi
\ifx \bpublisher  \undefined \def \bpublisher#1{#1}\fi
\ifx \bbtitle  \undefined \def \bbtitle#1{#1}\fi
\ifx \bedition  \undefined \def \bedition#1{#1}\fi
\ifx \bseriesno  \undefined \def \bseriesno#1{#1}\fi
\ifx \blocation  \undefined \def \blocation#1{#1}\fi
\ifx \bsertitle  \undefined \def \bsertitle#1{#1}\fi
\ifx \bsnm \undefined \def \bsnm#1{#1}\fi
\ifx \bsuffix \undefined \def \bsuffix#1{#1}\fi
\ifx \bparticle \undefined \def \bparticle#1{#1}\fi
\ifx \barticle \undefined \def \barticle#1{#1}\fi
\bibcommenthead
\ifx \bconfdate \undefined \def \bconfdate #1{#1}\fi
\ifx \botherref \undefined \def \botherref #1{#1}\fi
\ifx \url \undefined \def \url#1{\textsf{#1}}\fi
\ifx \bchapter \undefined \def \bchapter#1{#1}\fi
\ifx \bbook \undefined \def \bbook#1{#1}\fi
\ifx \bcomment \undefined \def \bcomment#1{#1}\fi
\ifx \oauthor \undefined \def \oauthor#1{#1}\fi
\ifx \citeauthoryear \undefined \def \citeauthoryear#1{#1}\fi
\ifx \endbibitem  \undefined \def \endbibitem {}\fi
\ifx \bconflocation  \undefined \def \bconflocation#1{#1}\fi
\ifx \arxivurl  \undefined \def \arxivurl#1{\textsf{#1}}\fi
\csname PreBibitemsHook\endcsname

\bibitem{Muller:2012zq}
\begin{barticle}
\bauthor{\bsnm{Muller}, \binits{B.}},
\bauthor{\bsnm{Schukraft}, \binits{J.}},
\bauthor{\bsnm{Wyslouch}, \binits{B.}}:
\batitle{{First Results from Pb+Pb collisions at the LHC}}.
\bjtitle{Ann. Rev. Nucl. Part. Sci.}
\bvolume{62},
\bfpage{361}--\blpage{386}
(\byear{2012})
{\href{https://arxiv.org/abs/1202.3233}{{arXiv:1202.3233}}}
{[hep-ex]}.
\doiurl{10.1146/annurev-nucl-102711-094910}
\end{barticle}
\endbibitem

\bibitem{CMS:2012krf}
\begin{barticle}
\bauthor{\bsnm{Chatrchyan}, \binits{S.}}, \betal:
\batitle{{Measurement of the pseudorapidity and centrality dependence of the
  transverse energy density in PbPb collisions at $\sqrt{s_{NN}}=2.76$ TeV}}.
\bjtitle{Phys. Rev. Lett.}
\bvolume{109},
\bfpage{152303}
(\byear{2012})
{\href{https://arxiv.org/abs/1205.2488}{{arXiv:1205.2488}}}
{[nucl-ex]}.
\doiurl{10.1103/PhysRevLett.109.152303}
\end{barticle}
\endbibitem

\bibitem{ALICE:2016igk}
\begin{barticle}
\bauthor{\bsnm{Adam}, \binits{J.}}, \betal:
\batitle{{Measurement of transverse energy at midrapidity in Pb-Pb collisions
  at $\sqrt{s_{\rm NN}} = 2.76$ TeV}}.
\bjtitle{Phys. Rev. C}
\bvolume{94}(\bissue{3}),
\bfpage{034903}
(\byear{2016})
{\href{https://arxiv.org/abs/1603.04775}{{arXiv:1603.04775}}}
{[nucl-ex]}.
\doiurl{10.1103/PhysRevC.94.034903}
\end{barticle}
\endbibitem

\bibitem{HotQCD:2014kol}
\begin{barticle}
\bauthor{\bsnm{Bazavov}, \binits{A.}}, \betal:
\batitle{{Equation of state in ( 2+1 )-flavor QCD}}.
\bjtitle{Phys. Rev. D}
\bvolume{90},
\bfpage{094503}
(\byear{2014})
{\href{https://arxiv.org/abs/1407.6387}{{arXiv:1407.6387}}}
{[hep-lat]}.
\doiurl{10.1103/PhysRevD.90.094503}
\end{barticle}
\endbibitem

\bibitem{STAR:2017sal}
\begin{barticle}
\bauthor{\bsnm{Adamczyk}, \binits{L.}}, \betal:
\batitle{{Bulk Properties of the Medium Produced in Relativistic Heavy-Ion
  Collisions from the Beam Energy Scan Program}}.
\bjtitle{Phys. Rev. C}
\bvolume{96}(\bissue{4}),
\bfpage{044904}
(\byear{2017})
{\href{https://arxiv.org/abs/1701.07065}{{arXiv:1701.07065}}}
{[nucl-ex]}.
\doiurl{10.1103/PhysRevC.96.044904}
\end{barticle}
\endbibitem

\bibitem{BRAHMS:2009wlg}
\begin{barticle}
\bauthor{\bsnm{Arsene}, \binits{I.C.}}, \betal:
\batitle{{Nuclear stopping and rapidity loss in Au+Au collisions at
  s(NN)**(1/2) = 62.4-GeV}}.
\bjtitle{Phys. Lett. B}
\bvolume{677},
\bfpage{267}--\blpage{271}
(\byear{2009})
{\href{https://arxiv.org/abs/0901.0872}{{arXiv:0901.0872}}}
{[nucl-ex]}.
\doiurl{10.1016/j.physletb.2009.05.049}
\end{barticle}
\endbibitem

\bibitem{BRAHMS:2003wwg}
\begin{barticle}
\bauthor{\bsnm{Bearden}, \binits{I.G.}}, \betal:
\batitle{{Nuclear stopping in Au + Au collisions at s(NN)**(1/2) = 200-GeV}}.
\bjtitle{Phys. Rev. Lett.}
\bvolume{93},
\bfpage{102301}
(\byear{2004})
{\href{https://arxiv.org/abs/nucl-ex/0312023}{{arXiv:nucl-ex/0312023}}}.
\doiurl{10.1103/PhysRevLett.93.102301}
\end{barticle}
\endbibitem

\bibitem{NA49:1998gaz}
\begin{barticle}
\bauthor{\bsnm{Appelshauser}, \binits{H.}}, \betal:
\batitle{{Baryon stopping and charged particle distributions in central Pb + Pb
  collisions at 158-GeV per nucleon}}.
\bjtitle{Phys. Rev. Lett.}
\bvolume{82},
\bfpage{2471}--\blpage{2475}
(\byear{1999})
{\href{https://arxiv.org/abs/nucl-ex/9810014}{{arXiv:nucl-ex/9810014}}}.
\doiurl{10.1103/PhysRevLett.82.2471}
\end{barticle}
\endbibitem

\bibitem{Mohs:2019iee}
\begin{barticle}
\bauthor{\bsnm{Mohs}, \binits{J.}},
\bauthor{\bsnm{Ryu}, \binits{S.}},
\bauthor{\bsnm{Elfner}, \binits{H.}}:
\batitle{{Particle Production via Strings and Baryon Stopping within a Hadronic
  Transport Approach}}.
\bjtitle{J. Phys. G}
\bvolume{47}(\bissue{6}),
\bfpage{065101}
(\byear{2020})
{\href{https://arxiv.org/abs/1909.05586}{{arXiv:1909.05586}}}
{[nucl-th]}.
\doiurl{10.1088/1361-6471/ab7bd1}
\end{barticle}
\endbibitem

\bibitem{Li:2018ini}
\begin{barticle}
\bauthor{\bsnm{Li}, \binits{M.}},
\bauthor{\bsnm{Kapusta}, \binits{J.I.}}:
\batitle{{Large Baryon Densities Achievable in High Energy Heavy Ion Collisions
  Outside the Central Rapidity Region}}.
\bjtitle{Phys. Rev. C}
\bvolume{99}(\bissue{1}),
\bfpage{014906}
(\byear{2019})
{\href{https://arxiv.org/abs/1808.05751}{{arXiv:1808.05751}}}
{[nucl-th]}.
\doiurl{10.1103/PhysRevC.99.014906}
\end{barticle}
\endbibitem

\bibitem{McLerran:2018avb}
\begin{barticle}
\bauthor{\bsnm{McLerran}, \binits{L.D.}},
\bauthor{\bsnm{Schlichting}, \binits{S.}},
\bauthor{\bsnm{Sen}, \binits{S.}}:
\batitle{{Spacetime picture of baryon stopping in the color-glass condensate}}.
\bjtitle{Phys. Rev. D}
\bvolume{99}(\bissue{7}),
\bfpage{074009}
(\byear{2019})
{\href{https://arxiv.org/abs/1811.04089}{{arXiv:1811.04089}}}
{[hep-ph]}.
\doiurl{10.1103/PhysRevD.99.074009}
\end{barticle}
\endbibitem

\bibitem{ALICE:2012ovd}
\begin{barticle}
\bauthor{\bsnm{Abelev}, \binits{B.}}, \betal:
\batitle{{Pion, Kaon, and Proton Production in Central Pb--Pb Collisions at
  $\sqrt{s_{NN}} = 2.76$ TeV}}.
\bjtitle{Phys. Rev. Lett.}
\bvolume{109},
\bfpage{252301}
(\byear{2012})
{\href{https://arxiv.org/abs/1208.1974}{{arXiv:1208.1974}}}
{[hep-ex]}.
\doiurl{10.1103/PhysRevLett.109.252301}
\end{barticle}
\endbibitem

\bibitem{ALICE:2013mez}
\begin{barticle}
\bauthor{\bsnm{Abelev}, \binits{B.}}, \betal:
\batitle{{Centrality dependence of $\pi$, K, p production in Pb-Pb collisions
  at $\sqrt{s_{NN}}$ = 2.76 TeV}}.
\bjtitle{Phys. Rev. C}
\bvolume{88},
\bfpage{044910}
(\byear{2013})
{\href{https://arxiv.org/abs/1303.0737}{{arXiv:1303.0737}}}
{[hep-ex]}.
\doiurl{10.1103/PhysRevC.88.044910}
\end{barticle}
\endbibitem

\bibitem{ALICE:2019hno}
\begin{barticle}
\bauthor{\bsnm{Acharya}, \binits{S.}}, \betal:
\batitle{{Production of charged pions, kaons, and (anti-)protons in Pb-Pb and
  inelastic $pp$ collisions at $\sqrt {s_{NN}}$ = 5.02 TeV}}.
\bjtitle{Phys. Rev. C}
\bvolume{101}(\bissue{4}),
\bfpage{044907}
(\byear{2020})
{\href{https://arxiv.org/abs/1910.07678}{{arXiv:1910.07678}}}
{[nucl-ex]}.
\doiurl{10.1103/PhysRevC.101.044907}
\end{barticle}
\endbibitem

\bibitem{Kapusta:1977zb}
\begin{barticle}
\bauthor{\bsnm{Kapusta}, \binits{J.I.}}:
\batitle{{Bremsstrahlung in the Nuclear Fireball Model}}.
\bjtitle{Phys. Rev. C}
\bvolume{15},
\bfpage{1580}--\blpage{1582}
(\byear{1977}).
\doiurl{10.1103/PhysRevC.15.1580}
\end{barticle}
\endbibitem

\bibitem{Bjorken:1984sp}
\begin{barticle}
\bauthor{\bsnm{Bjorken}, \binits{J.D.}},
\bauthor{\bsnm{McLerran}, \binits{L.D.}}:
\batitle{{Coherent Photon Radiation from Nuclei as a Probe of Impact Parameter
  and Nucleon Velocity Distribution in Ultrarelativistic Nuclear Collisions}}.
\bjtitle{Phys. Rev. D}
\bvolume{31},
\bfpage{63}
(\byear{1985}).
\doiurl{10.1103/PhysRevD.31.63}
\end{barticle}
\endbibitem

\bibitem{Dumitru:1993ph}
\begin{barticle}
\bauthor{\bsnm{Dumitru}, \binits{A.}},
\bauthor{\bsnm{McLerran}, \binits{L.D.}},
\bauthor{\bsnm{Stoecker}, \binits{H.}},
\bauthor{\bsnm{Greiner}, \binits{W.}}:
\batitle{{Soft photons at RHIC and LHC}}.
\bjtitle{Phys. Lett. B}
\bvolume{318},
\bfpage{583}--\blpage{586}
(\byear{1993}).
\doiurl{10.1016/0370-2693(93)90457-S}
\end{barticle}
\endbibitem

\bibitem{Jeon:1998tq}
\begin{barticle}
\bauthor{\bsnm{Jeon}, \binits{S.}},
\bauthor{\bsnm{Kapusta}, \binits{J.I.}},
\bauthor{\bsnm{Chikanian}, \binits{A.}},
\bauthor{\bsnm{Sandweiss}, \binits{J.}}:
\batitle{{Nucleus-nucleus bremsstrahlung from ultrarelativistic collisions}}.
\bjtitle{Phys. Rev. C}
\bvolume{58},
\bfpage{1666}--\blpage{1670}
(\byear{1998})
{\href{https://arxiv.org/abs/nucl-th/9806047}{{arXiv:nucl-th/9806047}}}.
\doiurl{10.1103/PhysRevC.58.1666}
\end{barticle}
\endbibitem

\bibitem{Kapusta:1999hb}
\begin{barticle}
\bauthor{\bsnm{Kapusta}, \binits{J.I.}},
\bauthor{\bsnm{Wong}, \binits{S.M.H.}}:
\batitle{{Imaging the space-time evolution of high-energy nucleus-nucleus
  collisions with bremsstrahlung}}.
\bjtitle{Phys. Rev. C}
\bvolume{59},
\bfpage{3317}
(\byear{1999})
{\href{https://arxiv.org/abs/hep-ph/9903235}{{arXiv:hep-ph/9903235}}}.
\doiurl{10.1103/PhysRevC.59.3317}
\end{barticle}
\endbibitem

\bibitem{Park:2021ljg}
\begin{barticle}
\bauthor{\bsnm{Park}, \binits{S.}},
\bauthor{\bsnm{Wiedemann}, \binits{U.A.}}:
\batitle{{Bremsstrahlung photons from stopping in heavy-ion collisions}}.
\bjtitle{Phys. Rev. C}
\bvolume{104}(\bissue{4}),
\bfpage{044903}
(\byear{2021})
{\href{https://arxiv.org/abs/2107.05129}{{arXiv:2107.05129}}}
{[hep-ph]}.
\doiurl{10.1103/PhysRevC.104.044903}
\end{barticle}
\endbibitem

\bibitem{Jackson1975}
\begin{bbook}
\bauthor{\bsnm{Jackson}, \binits{J.D.}}:
\bbtitle{Classical Electrodynamics}.
\bpublisher{Wiley},
\blocation{New York}
(\byear{1975})
\end{bbook}
\endbibitem

\bibitem{Park:2022kcs}
\begin{botherref}
\oauthor{\bsnm{Park}, \binits{S.}}:
{Opportunities with ultra-soft photons: Bremsstrahlung from stopping}
(2022)
{\href{https://arxiv.org/abs/2208.01527}{{arXiv:2208.01527}}}
{[hep-ph]}
\end{botherref}
\endbibitem

\bibitem{Bierlich:2022pfr}
\begin{botherref}
\oauthor{\bsnm{Bierlich}, \binits{C.}}, et al.:
{A comprehensive guide to the physics and usage of PYTHIA 8.3}
(2022)
{\href{https://arxiv.org/abs/2203.11601}{{arXiv:2203.11601}}}
{[hep-ph]}
\end{botherref}
\endbibitem

\bibitem{Klein:1999qj}
\begin{barticle}
\bauthor{\bsnm{Klein}, \binits{S.}},
\bauthor{\bsnm{Nystrand}, \binits{J.}}:
\batitle{{Exclusive vector meson production in relativistic heavy ion
  collisions}}.
\bjtitle{Phys. Rev. C}
\bvolume{60},
\bfpage{014903}
(\byear{1999})
{\href{https://arxiv.org/abs/hep-ph/9902259}{{arXiv:hep-ph/9902259}}}.
\doiurl{10.1103/PhysRevC.60.014903}
\end{barticle}
\endbibitem

\bibitem{Bierlich:2018xfw}
\begin{barticle}
\bauthor{\bsnm{Bierlich}, \binits{C.}},
\bauthor{\bsnm{Gustafson}, \binits{G.}},
\bauthor{\bsnm{L\"onnblad}, \binits{L.}},
\bauthor{\bsnm{Shah}, \binits{H.}}:
\batitle{{The Angantyr model for Heavy-Ion Collisions in PYTHIA8}}.
\bjtitle{JHEP}
\bvolume{10},
\bfpage{134}
(\byear{2018})
{\href{https://arxiv.org/abs/1806.10820}{{arXiv:1806.10820}}}
{[hep-ph]}.
\doiurl{10.1007/JHEP10(2018)134}
\end{barticle}
\endbibitem

\bibitem{Weil:2016zrk}
\begin{barticle}
\bauthor{\bsnm{Weil}, \binits{J.}}, \betal:
\batitle{{Particle production and equilibrium properties within a new hadron
  transport approach for heavy-ion collisions}}.
\bjtitle{Phys. Rev. C}
\bvolume{94}(\bissue{5}),
\bfpage{054905}
(\byear{2016})
{\href{https://arxiv.org/abs/1606.06642}{{arXiv:1606.06642}}}
{[nucl-th]}.
\doiurl{10.1103/PhysRevC.94.054905}
\end{barticle}
\endbibitem

\bibitem{LHCb:2014set}
\begin{barticle}
\bauthor{\bsnm{Aaij}, \binits{R.}}, \betal:
\batitle{{LHCb Detector Performance}}.
\bjtitle{Int. J. Mod. Phys. A}
\bvolume{30}(\bissue{07}),
\bfpage{1530022}
(\byear{2015})
{\href{https://arxiv.org/abs/1412.6352}{{arXiv:1412.6352}}}
{[hep-ex]}.
\doiurl{10.1142/S0217751X15300227}
\end{barticle}
\endbibitem

\bibitem{Klusek-Gawenda:2019ijn}
\begin{barticle}
\bauthor{\bsnm{K\l{}usek-Gawenda}, \binits{M.}},
\bauthor{\bsnm{McNulty}, \binits{R.}},
\bauthor{\bsnm{Schicker}, \binits{R.}},
\bauthor{\bsnm{Szczurek}, \binits{A.}}:
\batitle{{Light-by-light scattering in ultraperipheral heavy-ion collisions at
  low diphoton masses}}.
\bjtitle{Phys. Rev. D}
\bvolume{99}(\bissue{9}),
\bfpage{093013}
(\byear{2019})
{\href{https://arxiv.org/abs/1904.01243}{{arXiv:1904.01243}}}
{[hep-ph]}.
\doiurl{10.1103/PhysRevD.99.093013}
\end{barticle}
\endbibitem

\bibitem{ALICE:2020mso}
\begin{botherref}
\oauthor{\bsnm{ALICE}}:
{Letter of Intent: A Forward Calorimeter (FoCal) in the ALICE experiment}.
CERN-LHCC-2020-009, LHCC-I-036
(2020)
\end{botherref}
\endbibitem

\bibitem{ALICE:2803563}
\begin{botherref}
\oauthor{\bsnm{ALICE}}:
{Letter of intent for ALICE 3: A next generation heavy-ion experiment at the
  LHC},
Geneva.
CERN-LHCC-2022-009, LHCC-I-038
(2022)
\end{botherref}
\endbibitem

\end{thebibliography}

\end{document}